\documentclass[a4paper,fleqn,english]{article} 
\usepackage{latexsym}\usepackage{graphicx}\usepackage{babel}
\usepackage{epsfig}
\def\maketitle{\thispagestyle{empty}\setcounter{page}0\newpage
 \renewcommand{\thefootnote}{\arabic{footnote}}
 \setcounter{footnote}0}
\renewcommand{\thanks}[1]{\renewcommand{\thefootnote}{\fnsymbol{footnote}}
 \footnote{#1}\renewcommand{\thefootnote}{\arabic{footnote}}}

\renewcommand{\title}[1]{\begin{center}\Large\bf #1\end{center}\rm\par\bigskip}
\renewcommand{\author}[1]{\begin{center}\Large #1\end{center}}
\newcommand{\address}[1]{\begin{center}\large #1\end{center}}

\def\babs{\hrule\par\begin{description}\item{Abstract: }\it}
\def\eabs{\par\end{description}\hrule\par\medskip\rm}
\renewcommand{\date}[1]{\par\bigskip\par\sl\hfill #1\par\medskip\par\rm}

\begin{document}

\def\beq{\begin{equation}}
\def\eeq{\end{equation}}
\def\ber{\begin{eqnarray}}
\def\eer{\end{eqnarray}}
\def\l{\Lambda}
\def\lsim{\\lower-1.5pt\vbox{\hbox{\rlap{$<$}\lower5.3pt\vbox{\hbox{$\sim$}}}}\ }
\def\gsim{\
\lower-1.5pt\vbox{\hbox{\rlap{$>$}\lower5.3pt\vbox{\hbox{$\sim$}}}}\ }
\def\apj{{Astroph.\@ J.\ }}
\def\mn{{Mon.\@ Not.@ Roy.\@ Ast.\@ Soc.\ }}
\def\asta{{Astron.\@ Astrophys.\ }}
\def\aj{{Astron.\@ J.\ }}
\def\prl{{Phys.\@ Rev.\@ Lett.\ }}
\def\prd{{Phys.\@ Rev.\@ D\ }}
\def\nucp{{Nucl.\@ Phys.\ }}
\def\nat{{Nature\ }}
\def\plb {{Phys.\@ Lett.\@ B\ }}
\def \jetpl {JETP Lett.\ }
\def\ie {{i.e.}}
\def\n {\noindent}
\def\etal{{\it et al.}}
\def\m{{\rm m}}
\def\b{{\rm b}}
\def\d3h{\lower-1pt\hbox{$\stackrel{\ldots}{H}$}}
\def \ie {i.e.}
\def \lleq {\lower0.9ex\hbox{ $\buildrel < \over \sim$} ~}
\def \ggeq {\lower0.9ex\hbox{ $\buildrel > \over \sim$} ~}

\title{Modified gravity with vacuum polarization}

\author{Petr Tretyakov\thanks{tpv@xray.sai.msu.ru, tpv@theor.jinr.ru}}
\address{Joint Institute for Nuclear Research\\
Dubna, Russia}

\begin{abstract}
A brief review of cosmology in some generalized modified gravity theories with vacuum polarization is presented. Stability question of de Sitter solution is investigated.   
\end{abstract}

\section*{DGP brane dynamics with vacuum polarization}
The acceleration universe force us to explain this phenomena and this stimulate an interest to a number theories of modified gravity. One of such theories is the well-known Dvali-Gabadadze-Porrati (DGP) braneworld model \cite{DGP1,DGP2}, for instance, can lead to an accelerating universe without the presence of either a cosmological constant or some other form of dark energy. Generalizations of the DGP model can result in a phantom-like acceleration of the universe at late times, which is not excluded by observational data.We consider the simplest generic braneworld model with action of the form
\begin{equation}
\begin{array}{r}
S_{DGP}=M^3\left[ \int_{bulk}(R_5 -2\Lambda)\sqrt{-G}d^5x
-2\int_{brane} K\sqrt{-g}d^4x \right]
\\
\\
+\int_{brane}(m^2R_4-2\lambda)\sqrt{-g}d^4x
+\int_{brane}L(g_{ab},\rho,\phi)\sqrt{-g}d^4x.
\end{array}
\label{ss01}
\end{equation}
Here, $R_5$ is the scalar curvature of the metric $G_{ab}$ in the
five-dimensional bulk, and $R_4$ is the scalar curvature of the induced metric
$g_{ab} = G_{ab} - n_a n_b$ on the brane, where $n^a$ is the vector field of
the inner unit normal to the brane, which is assumed to be a boundary of the
bulk space, and the notation and conventions of \cite{Wald} are used. The
quantity $K = g^{ab} K_{ab}$ is the trace of the symmetric tensor of extrinsic
curvature $K_{ab} = g^c{}_a \nabla_c n_b$ of the brane. The symbol $L (g_{ab},
\phi)$ denotes the Lagrangian density of the four-dimensional matter fields
$\phi$ whose dynamics is restricted to the brane so that they interact only
with the induced metric $g_{ab}$. All integrations over the bulk and brane are
taken with the corresponding natural volume elements. The symbols $M$ and $m$
denote the five-dimensional and four-dimensional Planck masses, respectively,
$\Lambda$ is the bulk cosmological constant, and $\lambda$ is the brane
tension (it may be interpreted as cosmological constant on the brane). Note also that original DGP model apply $\lambda=0$.

In $Z_2$ symmetry case cosmological constant is equal from two side of the brane $\Lambda_1=\Lambda_2=\Lambda$, and
corresponding dynamical equation in FRW background looks like \cite{Collins1,Shtanov}:
\begin{equation}
H^2+\frac{k}{a^2}=\frac{\rho+\lambda}{3m^2}+\frac{2}{l^2}\left[1\pm
\sqrt{1+l^2\left(\frac{\rho+\lambda}{3m^2}
-\frac{\Lambda}{6}-\frac{C}{a^4} \right) } \right].
\label{37}
\end{equation}

Here $l=2m^2/M^3$, $H \equiv \dot a/a$ is the Hubble parameter, and $\rho$ is the matter
energy density on the brane. Here and below, the overdot derivative is taken
with respect to the cosmological time $t$ on the brane. The expression under square root tend to zero during the evolution. It mean that in some time all time derivative from Habble parameter $\dot H,
\ddot H, ...$ tends to infinity while $H$ is finite. It's a new specific type of singularity and it was studied very well in the literature \cite{Shtanov1}.
The term containing the constant $C$ describes the so-called ``dark
radiation.'' We don't take into account this term, also
we consider a spatially flat universe ($k = 0$).
The ``$\pm$'' signs in the solution correspond to two branches defined by the
two possible ways of bounding the Schwarzschild--(anti)-de~Sitter bulk space by
the brane \cite{CH,Deffayet}.
\begin{figure}
\begin{center}
\includegraphics[width=0.45\textwidth]{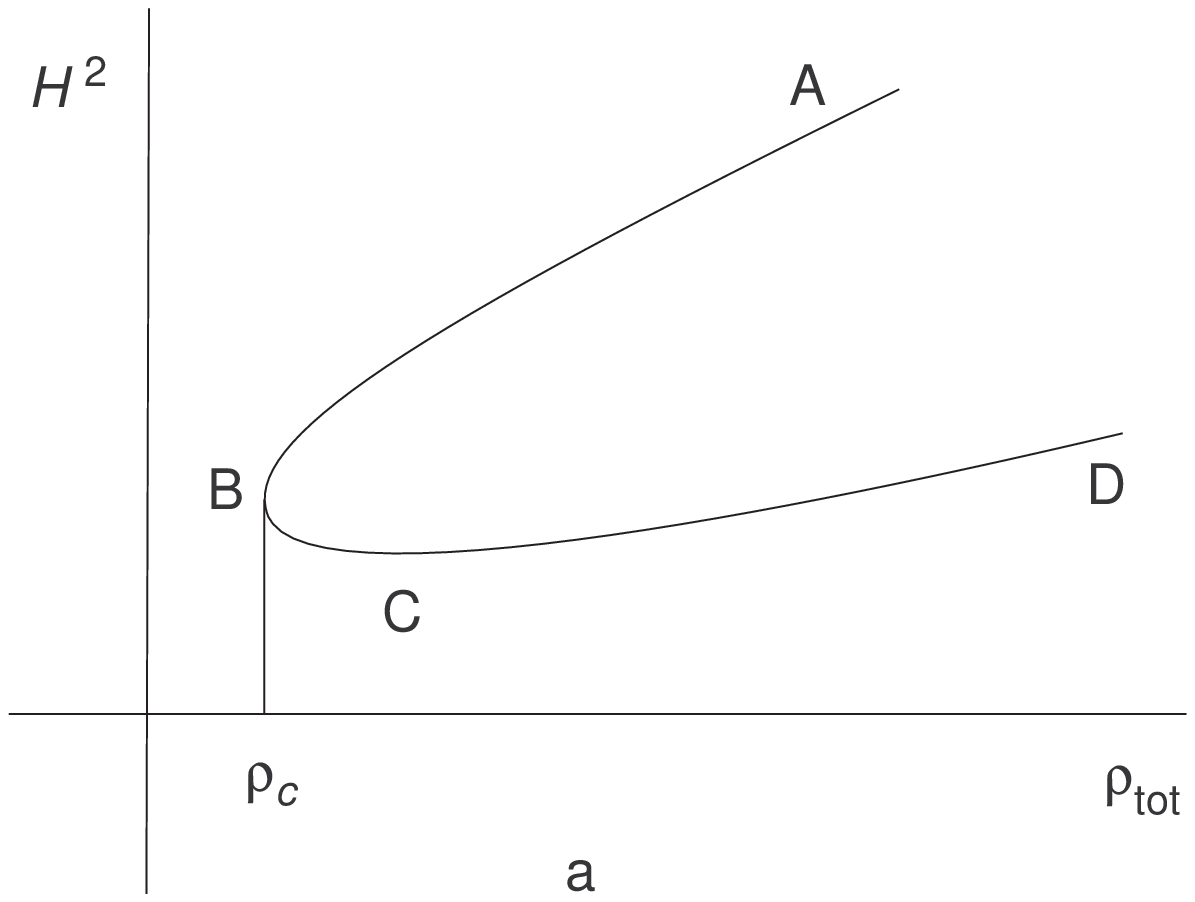} \hspace{1cm}
\includegraphics[width=0.45\textwidth]{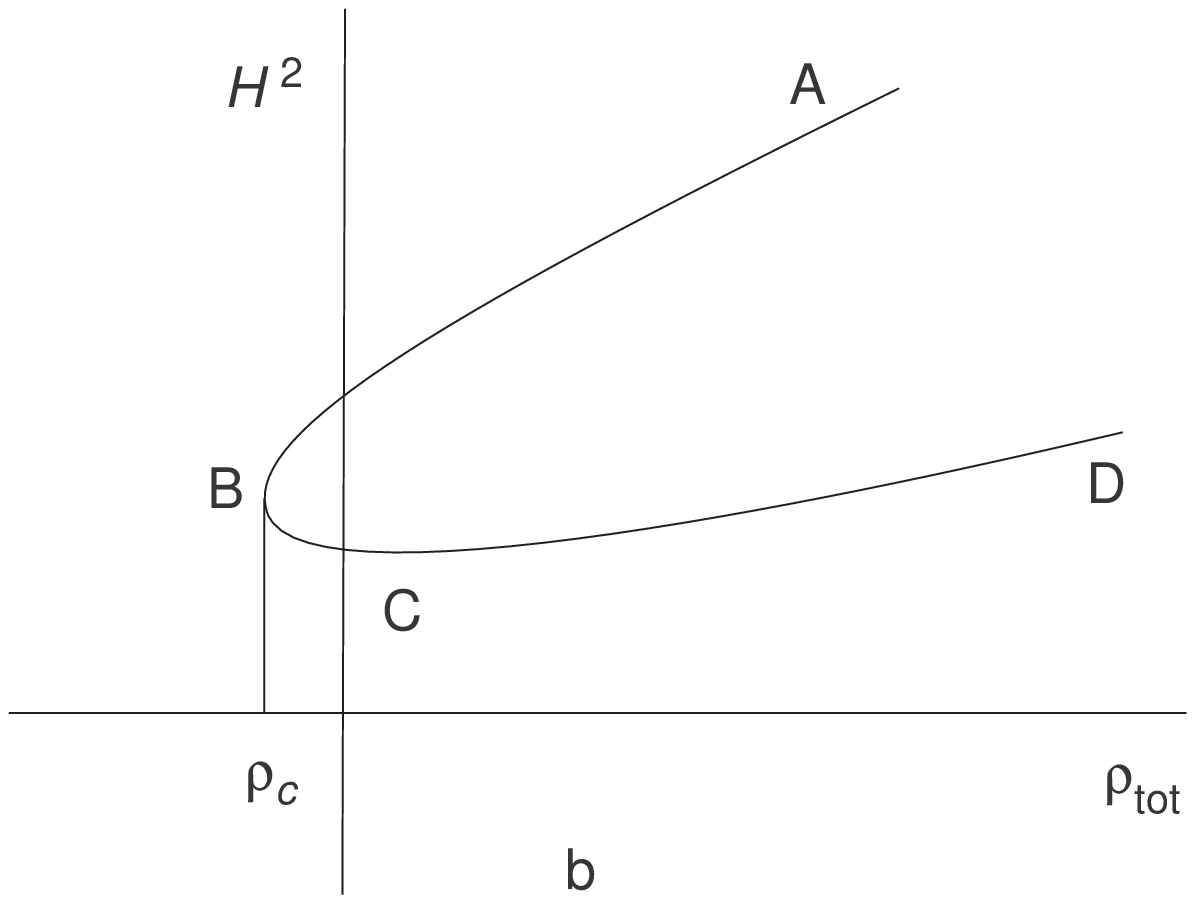}
\end{center}
\caption{Plot of relation (\ref{37}) in the case $\Lambda
> 0$. Case (a) corresponds to $\rho_c > 0$, and case (b) corresponds to $\rho_c
< 0$, where $\rho_c\equiv 3 m^2 \left(\Lambda/6 -
l^{-2}\right)$.}
\label{pos}
\end{figure}

Now let us briefly consider dynamic of the DGP brane. Classical dynamics depends significantly on the bulk cosmological constant $\Lambda$ and the full picture is sufficiently complicate so we refer you to the original paper focusing on only essential point for further statement ($\Lambda>0$).
The case $\Lambda>0$ is shown in Fig.\ref{pos}. The graph of
(\ref{37}) in the $\left( H^2, \rho_{\rm tot} \right)$ plane in
Fig.\ref{pos} illustrates that in an expanding universe the matter density
$\rho$ decreases (except for a ``phantom matter'' which we do not consider in
the present paper), and the point in the plane $(H^2,\rho_{\rm tot})$ moves
from right to left in Fig.\ref{pos}.

A striking feature of Fig.~\ref{pos} is that the value of the Hubble parameter
in the braneworld can {\em never drop to zero}. In other words, the Friedmann
asymptote $H \to 0$ is absent in our case. The upper and lower branches in
Fig.~\ref{pos} describe the two complementary braneworld models: branches AB
and DB are associated with Brane\,2 and Brane\,1 of \cite{SS}, respectively,
while branches AC and DC correspond to the lower and upper signs in
(\ref{37}), respectively, and describe the two branches with different
embedding in the bulk. It should be noted that, in many important cases, the
behaviour of the braneworld does not have any parallel in conventional
Friedmannian dynamics (by this we mean standard GR in a FRW universe). For
instance, the BC part of the evolutionary track corresponds to ``phantom-like''
cosmology with $\dot H>0$, even though matter on the brane never violates the
weak energy condition.

Now let us consider a some quantum effects. In general, quantum effects in curved space-time can arise on account of the
vacuum polarization as well as particle production. It is well known that the
latter is absent for conformally invariant fields (which we shall consider here) and that, in this case, quantum corrections to the equations of
motion are fully described by the renormalized vacuum energy--momentum tensor
which has the form \cite{BD}
\begin{equation}
\begin{array}{l}
\langle
T_{ik}\rangle\!=\!(m_2/2880\pi^2)(R_i^{\,\,l}R_{kl}-\frac{3}{2}RR_{ik}
-\frac{1}{2}g_{ik}R_{lm}R^{lm}+\frac{1}{4}g_{ik}R^2)\\
\\
\,\,\,\,\,\,\,\,+(m_3/2880\pi^2)\frac{1}{6}(2R_{\,;i;k}-2g_{ik}R^{\,;l}_{\,\,\,;l}
-2RR_{ik}+\frac{1}{2}g_{ik}R^2),
\end{array}
\label{35}
\end{equation}
where $m_1, m_2$ depend upon the spin weights of the different
fields contributing to the vacuum polarization. This effect is known for a long time in cosmology. For example, it was demonstrate the possibility of singularity problem solution by some specific changes of $m_2$ and $m_3$ \cite{Starobinsky}.

Since we work in flat FRW space-time, it is comfortable to use next relation:

\begin{equation}
\rho_q=k_2H^4+k_3(2\ddot H H+6\dot H H^2 -\dot H^2). \label{kvmat}
\end{equation}
Where parameters $k_2$ and $k_3$ take the next form
\cite{Odintsov1,Odintsov2,Odintsov3}:

\begin{equation}
k_2=\frac{m_2}{60(4\pi)^2}=\frac{N+11N_{1/2}+62N_1+1411N_2-28N_{HD}}{60(4\pi)^2},
 \label{kvmat1}
\end{equation}

\begin{equation}
k_3=\frac{m_3}{60(4\pi)^2}=\frac{N+6N_{1/2}+12N_1+611N_2-8N_{HD}}{60(4\pi)^2}.
 \label{kvmat2}
\end{equation}
Here $N_i$ is number of the fields with spin $i$ contributing to the vacuum polarization:
$N$ -- a number of scalar fields,  $N_{1/2}$
--  a number of fermion fields, $N_1$ -- a number of vector fields, $N_2$ ($=0$
or $1$) -- a number of gravitons, $N_{HD}$ -- a number of conformal scalar fields.
Note also that there is strong restriction on $k_2$ and $k_3$ in usual 4D-space-time. At least need $k_3<0$ with any $k_2$ for de Sitter (and in particular Minkovsky!) solution stability. Since we see that Minkovsky solution is stable (there isn't particle production in conformal flat vacuum) we may use condition $k_3<0$ as observational data.

In order to assess the effects of the vacuum polarization on the dynamics of
the braneworld, one must add $\rho_q$ to the matter density in (\ref{37}) so that $\rho \to \rho+\rho_q$ in those
equations. An important consequence of this operation is that the form of the
equation of motion changes dramatically\,---\,the original algebraic equation
changes to a differential equation\,! The dynamical equation (\ref{37}) now
takes the form
\begin{equation}\label{dyn}
\ddot H H \!\!=\!\! \frac12 \dot H^2 -3\dot H H^2 + \left( 2k_3 \right)^{-1}\!\! \left(
\!-\!k_2 H^4\!+\!3 m^2 H^2\!-\!\rho_{\rm tot}\! \pm\! 3 M^3 \sqrt{H^2 \!-\!\Lambda/6}
\right)  \, .
\end{equation}

The goal of the present paper is to study the stability of the classical
solutions when vacuum polarization terms are taken into account. The $k_2$-term
in (\ref{kvmat}), which does not contain time derivatives of $H$, can only
change the position of the future stable points. On the contrary, due to the
$k_3$-term in (\ref{kvmat}), some classical solutions can lose stability.
Therefore, for simplicity (and without loss of generality), we set $k_2=0$ in
our calculations. If the brane has nonzero tension $\lambda$,
the stationary points of (\ref{dyn}) in the case $k_2=0$ can be found by
substituting $\lambda$ into (\ref{37}) and setting $\rho=0$. After that, we
linearize Eq.~(\ref{dyn}) at these stationary points and find the eigenvalues
of the corresponding linearized system. The condition of stability of the
stationary point is that its eigenvalues's real parts are negative.

The eigenvalues at the stationary points where $\dot H=0$ are given by
\begin{equation}
\mu_{1,2}=\frac{1}{2} \left( f_1 \pm \sqrt{f_1^2+4f_2} \right) \, , \label{mu}
\end{equation}
where we have made the notation
\begin{equation}
f_1 = -3H \, ,
\end{equation}
\begin{equation}
f_2 = \frac{1}{2k_3} \left(1+\frac{\lambda}{3 m^2H^2} \pm \frac{2
l^{-1}\Lambda/6} {H^2 \sqrt{H^2-\Lambda/6}}\right)\, , \label{f_2}
\end{equation}

Two different signs in Eq.~(\ref{f_2}) correspond to two different equations of
motion, while, in Eq.~(\ref {mu}), we have two different eigenvalues of a
single equation.

Since $f_1$ is negative, the eigenvalue $\mu_2$ corresponding to the ``$-$''
sign in Eq.~(\ref{mu}) is also always negative. Moreover, $\mu_1$ is positive
if and only if $f_2$ is positive. As a result, the stability of a fixed point
is equivalent to the condition $f_2 < 0$. More careful investigation of this condition (for details see original paper \cite{TTSS}) show that part BC on the Fig.~\ref{pos} corresponding to effective phantom behavior is unstable with respect vacuum polarization.

\section*{General brane dynamics with vacuum polarization}

Now let us generalize obtaining result. We investigate any theory, which lead to dynamical equation in the next form:
\begin{equation}
\rho=F(H),
 \label{186}
\end{equation}
where $F(H)$ may be any algebraical function, which is don't contain time derivative of $H$. First of all we investigate the case $k_2=0$. Substituting in (\ref{186}) expression (\ref{kvmat}) for $\rho_q$ we may rewrite equation as dynamical system:
\begin{equation}
\begin{array}{l}
\dot H=C,\\
\dot C=\frac{C^2}{2H}-3CH+\frac{1}{2k_3H}F(H)\equiv f(H,C).
\end{array}
 \label{187}
\end{equation}
Linearizing this system at the fixed point $(H_0,0)$
\begin{equation}
\begin{array}{l}
\dot H=C,\\
\dot C=(\frac{\partial f}{\partial C})_0C+(\frac{\partial
f}{\partial H})_0H,
\end{array}
\label{173}
\end{equation}
we find its eigenvalues
\begin{equation}
 \mu_{1,2}=\frac{1}{2}[(\frac{\partial f}{\partial C})_0\pm \sqrt{(\frac{\partial f}{\partial C})_0^2 +4(\frac{\partial
f}{\partial H})_0} ].\label{174}
\end{equation}
Since $(\frac{\partial f}{\partial C})_0=-3H_0<0$ the real part of eigenvalues $\mu_{1,2}$ takes negative values if and only if

\begin{equation}
\left( \frac{\partial f}{\partial H}
\right)_0=\frac{1}{2H_0k_3}\left(\frac{\partial F}{\partial
H}\right)_0<0.
 \label{188}
\end{equation}
So we can see that it need condition $(\frac{\partial F}{\partial
H})_0>0$ for stability. From another hand using (\ref{186}) we find $(\frac{\partial F}{\partial
H})_0=\frac{\partial \rho}{\partial H}>0$ or in this case the equivalent condition $\frac{\partial H}{\partial \rho}>0$.
By another words all regimes with $\frac{\partial \rho}{\partial H}<0$ is unstable with respect to vacuum polarization.

Rewriting last condition in the form $\frac{\partial \rho /
\partial t}{\partial H / \partial t}<0$ we find that any regime with $\dot \rho<0$ (that is natural in expanding universe) and with $\dot H>0$ is unstable. Exactly similar regimes is called fantom-like. The case describing here correspond for example to BC part in Fig.~\ref{pos}.

Now let us account the $k_2$ term contribute. It mean that a new term $k_2H^4$ is appear in equation (\ref{186}). Transferring this term into right hand part of equation we may introduce a new function $F^{\prime}(H)=F(H)-k_2H^4$ and all previous result is true for this function. So term $k_2H^4$ is not influencing on the stability, but it may change dynamical equation and fantom regimes is disappear at all \cite{TT}.

\section*{$f(R)$-theories with vacuum polarization}

 Now let us consider $f(R)$-theories which more popular in resent time. The most general form of such theories may be written in the next form\cite{Starobinsky1} (for a more general review of $f(R)$ gravity, see aslo \cite{Odintsov4,Odintsov5}):
\begin{equation}
S=S_m+\frac{1}{2\chi}\int d^4x\sqrt{-g}f(R),
\label{3-01}
\end{equation}
here $\chi=8\pi G$ and for the sake of simplicity we set $2\chi =1$. The general equation of motion corresponding to (\ref{3-01}) is given by

\begin{equation}
g_{ik}f'^{;l}_{;l}-f'_{;i;k}+f'R_{ik}-\frac{1}{2}fg_{ik}=T_{ik}.
\label{3-02}
\end{equation}
The trace of equation of motion (\ref{3-02}) reads
\begin{equation}
3f'^{;l}_{;l}+f'R-2f=T.
\label{3-03}
\end{equation}
So existence condition of de Sitter solution in vacuum is \cite{Cognola, Tretyakov}
\begin{equation}
2f(R_0)-R_0f'(R_0)=0,
\label{3-04}
\end{equation}
and condition of its stability following from (\ref{3-03}) is given by
\begin{equation}
\frac{f'(R_0)}{R_0f''(R_0)}-1>0.
\label{3-05}
\end{equation}
Now let us take into account the vacuum polarization effects. For the sake of simplicity we investigate only $k_2=0$ case. So we find from (\ref{35}) $\langle T_{l}^l\rangle =-\frac{k_3}{3}R^{;l}_{;l}$. Substituting this relation into (\ref{3-03}) we find that this equation is restored to a normal state (without $\langle T_{l}^l\rangle$ contribute) by the $f\rightarrow f-\frac{k_3}{18}R^2$ transformation. And substituting this one into (\ref{3-05}) we find new de Sitter stability condition:
\begin{equation}
\frac{f'(R_0)-R_0f''(R_0)}{R_0(f''(R_0)-\frac{k_3}{9})}>0,
\label{3-06}
\end{equation}
which is turn into the expression (\ref{3-05}) by the limit $k_3=0$.
For instance, let us consider meaning of new expression (\ref{3-06})  for the model \cite{Odintsov6}
$f(R)=R+R^{-m}+R^n$. We see that only denominator of the (\ref{3-06}) has change when vacuum polarization take into account, while numerator is unchangeable. From another hand for this class of theories $f''=m(m+1)R^{-m-2}+n(n+1)R^{n-2}>0$ and since $k_3<0$ taking into account of vacuum polarization effect can't change denominator's sign. It mean that in this case quantum effects do not influence on de Sitter stability condition. 
Note also that vacuum polarization effects to above $f(R)$ gravity model may
suppress instabilities as it was noted in \cite{Odintsov7}.

\section*{Conclusion} 
Some modified gravity theories was studied with respect to vacuum polarization. It was demonstrate instability of effective fantom regimes in the brane cosmology caused by vacuum polarization. Modified condition of the vacuum de Sitter solution stability in $f(R)$ theories has been derived.

\section*{Acknowledgments} 
This is work based mainly on results obtained in collaboration with A. Toporensky, V. Sahni, Yu. Shtanov and i would like to thank them. I would like to thank S.D. Odintsov for some useful discussions. 
This work was partially supported by RFBR grant 08-02-00923 and
with the scientific school grant 4899.2008.2 of the Russian Ministry of Science and Technology.

\end{document}